\documentclass[trackchanges,twocolumn]{aastex701}

\usepackage{braket}
\usepackage{amsthm}
\usepackage{ascmac}
\usepackage{amssymb, amsmath}
\usepackage{bm}
\usepackage{physics}
\usepackage{graphicx}
\usepackage{booktabs}

\usepackage{times}
\usepackage[T1]{fontenc}
\usepackage{color}
\usepackage{hyperref}
\usepackage{booktabs}
\usepackage{multirow}

\begin{document}
\title{Accuracy and Applicability of the Hartle–Thorne and Komatsu-Eriguchi-Hachisu Methods \\
for Modeling Rotating Neutron Stars}

\author[orcid=0009-0003-2023-4146,sname='Kwon']{Hyukjin Kwon}
\affiliation{Department of Physics, School of Science, Institute of Science Tokyo, Tokyo 152-8550, Japan}
\email{kwon.h.04c4@m.isct.ac.jp}

\author[orcid=0009-0001-8708-5486,sname='Yoshimura']{Kenta Yoshimura}
\affiliation{Department of Physics, School of Science, Institute of Science Tokyo, Tokyo 152-8550, Japan}
\email{yoshimura.k.af21@m.isct.ac.jp}

\author[orcid=0000-0001-9186-8793,sname='Miyatsu']{Tsuyoshi Miyatsu}
\affiliation{Department of Physics and Origin of Matter and Evolution of Galaxies Institute, Soongsil University, Seoul 06978, Republic of Korea}
\email{tsuyoshi.miyatsu@ssu.ac.kr}

\author[orcid=0000-0001-5800-1995,sname='Sekizawa']{Kazuyuki Sekizawa}
\affiliation{Department of Physics, School of Science, Institute of Science Tokyo, Tokyo 152-8550, Japan}
\affiliation{Division of Nuclear Physics, Center for Computational Sciences, University of Tsukuba, Ibaraki 305-8577, Japan}
\affiliation{RIKEN Nishina Center, Saitama 351-0198, Japan}
\email{sekizawa@phys.sci.isct.ac.jp}

\author[orcid=0000-0001-7810-5134,sname='Cheoun']{Myung-Ki Cheoun}
\affiliation{Department of Physics and Origin of Matter and Evolution of Galaxies Institute, Soongsil University, Seoul 06978, Republic of Korea}
\email{}

\begin{abstract}
    Neutron stars, which are composed of extremely dense nuclear matter, serve as natural laboratories to study nuclear interactions beyond the terrestrial experiments. Recent studies have actively explored how the equation of state (EoS) can be constrained by observed neutron star masses and radii, and how nuclear interactions affect their macroscopic properties. Most of these studies, however, rely on the Tolman-Oppenheimer-Volkoff (TOV) equations, which assumed static, spherically symmetric neutron stars. Since neutron stars are rotating objects and thus axisymmetrically deformed, the TOV calculation may be insufficient to capture their realistic structure. In this work, we investigate the influence of nuclear matter properties on the physical quantities of rotating neutron stars using two approaches: the perturbative Hartle-Thorne (HT) method and fully general relativistic Komatsu-Eriguchi-Hachisu (KEH) method.
    For nuclear EoS parameter sets, we examine the OMEG series, in which the slope of the symmetry energy $L$ is systematically varied.
    We find that rotational effects lead to a noticeable increase in the stellar radius, which depends sensitively on values of $L$. 
    Additionally, focusing on the rotational deformation, we show that the results obtained by these two methods deviate from each other even for the slowly rotating case such as $f=200$ Hz.
    These results reveal that, for detailed discussions on the internal structure and stability of rotating neutron stars, the fully general relativistic method such as KEH is indispensable.
\end{abstract}


\section{Introduction}
Neutron stars are compact objects whose structure should be described within the framework of the  general relativity. The first theoretical attempt to describe the hydrodynamical balance between the pressure and gravity under Einstein’s field equations \citep{Einstein1916} was made by \cite{Tolman1934,Oppenheimer1939}, resulting in what is now known as the Tolman-Oppenheimer-Volkoff (TOV) equation. The TOV equation consists of a couple of first-order ordinary differential equations (ODEs), which can be easily solved using numerical integrations. Owing to its simplicity and physical relevance, it has been widely applied not only in astrophysics but also in nuclear and particle physics communities.

However, the TOV framework assumes the static and spherically symmetric configuration, which limits its applicability to non-rotating neutron stars. 
On the other hand, observational evidence indicates that neutron stars are indeed rotating \citep{Gotthelf2013}, with some exhibiting extremely rapid spin rates. Notable examples include PSR J0030+0451 (205 Hz) \citep{Miller2019}, PSR J0740+6620 (346 Hz) \citep{Wolff2021}, and PSR J1748–2446ad (716 Hz) \citep{Hessels2006}, which rotate with millisecond periods and are thus classified as millisecond pulsars. For such rapidly rotating neutron stars (RNSs), rotational effects become non-negligible; nevertheless, there have been a limited number of studies taking into account these rotational effects.

To model slowly rotating neutron stars, the works by \cite{Hartle1967} and \cite{Hartle1968} developed a perturbative approach that allows the stellar structure to be calculated using an equation of state (EoS) table based on simplified models \citep{Harrison1965,Tsuruta1966}. This Hartle-Thorne (HT) method yields a coupled set of first-order ODEs and provides perturbative results starting from the TOV solution. In \cite{weber1992}, this method was implemented for various EoSs to facilitate further discussion based on realistic EoSs.
However, the HT method is applicable only to slowly rotating configurations and cannot fully capture relativistic effects at higher spin rates. To model RNSs more accurately, additional centrifugal-force terms induced by rotation must be included. These effects deform the stellar shape, transforming the originally one-dimensional hydrostatic problem into a two-dimensional one with axial-symmetry that depends on both the radius and the polar angle.

Several numerical approaches have been developed to address this point. A detailed theoretical overview of these numerical approaches can be found in \cite{Paschalidis2017}. Among them, the Hachisu Self-Consistent Field (HSCF) method \citep{Hachisu1986} remains one of the most stable approaches. The HSCF method expresses the equilibrium condition among pressure, gravity, and centrifugal force in an integral form of the Poisson equations, enabling self-consistent computation. This approach was later extended to the general relativitic formalism, resulting in the Komatsu-Eriguchi-Hachisu (KEH) method \citep{Komatsu1989a,Komatsu1989b}. In the KEH framework, the Einstein field equations are reformulated into Poisson-like integral equations. Although the original KEH formulation had a problem that the computational domain of the radius is defined over $0<r<\infty$, \cite{Cook1992,Cook1994} later mapped the infinite domain to a finite interval $0<s<1$ by introducing a new radial coordinate $s$.
As implementation examples of the KEH method, Cook-Shapiro-Teukolsky (CST) approach has successfully modeled rapidly rotating neutron stars, employing a polytropic EoS. Subsequently,  \cite{Stergioulas1995} developed the \texttt{RNS} code, making these calculations applicable to tabulated EoSs and releasing it as open source software. The \texttt{RNS} code was later modified by Morsink to incorporate additional features in accordance with their works \citep{Laarakkers1999,Morsink1999}.

Another well-known method is the Bonazzola-Gourgoulhon-Salgado-Marck (BGSM) method, which solves elliptic differential equations using a spectral method. They also provided their code, \texttt{LORENE}, as open-source software \citep{Bonazzola1993, Bonazzola1998}. The BGSM method was applied to various EoSs in \cite{Salgado1994}, and the results showed excellent agreement with those of CST \citep{Cook1994}. A direct comparison of the two methods, KEH and BGSM, was conducted by \cite{Nozawa1998}. They showed that the differences between the two methods for the same EoS are not significant, even though the stiffness of the EoS contributes to the discrepancy.

However, there have been few examples where the discrepancy between the perturbative HT method and these full-relativistic methods were systematically discussed. Although \cite{Berti2004, Berti2005} performed a comparison between the HT and KEH methods and the analytic exterior solution of \cite{Manko2000a,Manko2000b}, 
based on quadrupole moments, 
the discussion was largely restricted to the exterior spacetime, 
with limited attention given to the internal structure of neutron stars.

For discussing the inner structure of neutron stars, it is also indispensable to model the pressure at each position, informed from the nuclear and particle physics.
The EoS, formulated by the pressure as a function of the baryon density, constitutes an important research topic in nuclear physics.
Neutron stars contain nuclear matter in a wide density region up to nearly ten times of the nuclear saturation density. 
In terrestrial laboratories, however, nuclear matter can be probed only near the saturation density. 
The EoSs of nuclear matter are characterized by values such as incompressibility $(K_0)$, symmetry energy $(S)$, and the slope of the symmetry energy $(L)$, which have been the focus of intensive theoretical and experimental studies. 
Recent works have reported that these values are correlated with macroscopic neutron-star properties. 
In particular, \cite{Sun2024} has demonstrated a strong correlation between the $L$ and neutron star radii using both non-relativistic Skyrme \citep{Skyrme1958}, Gogny \citep{Gogny1980} type, and relativistic mean field (RMF) models~\citep{Walecka1974}. 

The $L$ has been also closely related to the radii of RNSs.
For instance, \cite{Lopes2024} has reported that the $L$ affects not only the radii of a neutron stars with masses of $M=1.4M_\odot$ but also the enhancement of radius due to their rotation using the RMF framework.
However, that work has employed the HT method which is valid only for slowly RNSs, though it has been applied to the frequency of $800$ Hz.
\cite{Kacskovics2022} demonstrated that the accuracy of the HT approach deteriorates rapidly above $400$ Hz when compared with fully relativistic calculations.
This is why, to obtain a more reliable assessment of the correlation between rotation and $L$, fully general relativistic models of rotating neutron stars are required. 
Our previous work \citep{Kwon2025} has performed  calculations based on the KEH method using five different Skyrme-type EoSs. 
Our results have shown that, for a canonical $1.4 M_\odot$ neutron star, models with smaller $L$ values can sustain higher rotation frequencies, while those with larger $L$ values exhibit greater radius increases due to rotation. However, since the nuclear matter EoS that determines the structure of neutron stars depends on various nuclear matter properties, a more sophisticated treatment is required to evaluate the pure effect of $L$.

In the present study, our objectives are twofold: the first is to critically assess the applicability and limitations of the HT perturbative approach by direct comparisons with the KEH method,
and the second is to explore how the $L$ affects the rotational deformation and radius variation of neutron stars. 
For this purpose, we adopt the series of OMEG EoS, which has been developed within the RMF framework incorporating non-linear couplings between isoscalar and isovector mesons as well as the contributions of $\delta$ meson~\citep{Miyatsu2023}.
Its model parameters are calibrated to simultaneously satisfy constraints from heavy-ion collision flow data, precise neutron-star mass measurements (PSR J0740+6620), tidal deformability from GW170817~\citep{Abbott2018}, and the PREX-2 and CREX experiments.~\citep{Adhikari2021,Adhikari2022}
Among the OMEG series, all parameter sets have the same key nuclear matter properties such as saturation density $\rho_0$, effective nucleon mass $M^*$ and $K_0$, while only $L$ varies.
Thereby we can systematically investigate the net correlation between $L$ and neutron-star radii.
Inputting these EoSs, we calculate the mass-radius relations and other properties using both the HT and KEH methods.

The arrangement of this article is as follows.
Section 2 provides the brief formalism of our calculations.
Section 3 shows the calculation results and discusses them.
Section 4 summarizes the point of this article and mentions the prospects.

\section{Theoretical Framework}

\subsection{Nuclear Matter Equation of States}

The RMF model was originally formulated within the framework of quantum hydrodynamics (QHD) with two mesons, $\sigma$ and $\omega^\mu$~\citep{Walecka1974,Serot1984}.
This model was later extended to include two more mesons, $\boldsymbol{\delta}$ and $\boldsymbol{\rho}^\mu$, as well as their non-linear coupling terms.
The generic Lagrangian density in the RMF model is expressed as
\begin{equation}
\begin{aligned}
    \mathcal{L} &=\bar{\psi}_N[i\gamma_\mu \partial^\mu-(M_N-g_\sigma \sigma - g_\delta \boldsymbol{\delta}\cdot \boldsymbol{\tau}_N ) \\
    &\quad -g_\omega\gamma_\mu \omega^\mu - g_\rho \gamma_\mu \boldsymbol{\rho}^\mu \cdot \boldsymbol{\tau}_N]\psi_N \\
    &\quad + \frac{1}{2} (\partial_\mu \sigma \partial^\mu \sigma - m^2_\sigma \sigma^2) + \frac{1}{2}m^2_\omega \omega_\mu \omega^\mu - \frac{1}{4} W_{\mu \nu} W^{\mu \nu} \\
    &\quad + \frac{1}{2}(\partial_\mu \boldsymbol{\delta} \cdot \partial^\mu \boldsymbol{\delta} - m^2_\delta \boldsymbol{\delta} \cdot \boldsymbol{\delta}) +\frac{1}{2} m^2_\rho \boldsymbol{\rho}_\mu \cdot \boldsymbol{\rho}^\mu \\
    &- \frac{1}{4} \boldsymbol{R}_{\mu \nu} \cdot \boldsymbol{R}^{\mu \nu}
     - U_{\text{NL}}(\sigma,\omega,\boldsymbol{\delta},\boldsymbol{\rho}),
\end{aligned}
\end{equation}
where $\psi_N$ is the nucleon field, $\boldsymbol{\tau}_N$ denotes the isospin Pauli matrices, and $W_{\mu \nu} = \partial_\mu \omega_\nu - \partial_\nu \omega_\mu$ and $\boldsymbol{R}_{\mu \nu}  = \partial_\mu \boldsymbol{\rho}_\nu - \partial_\nu \boldsymbol{\rho}_\mu$ represent the field tensors for the $\omega$ and $\boldsymbol{\rho}^\mu$ mesons, respectively. The coupling constants are denoted by $g_i$ ($i=\{\sigma,\delta,\omega,\rho\}$). The nonlinear potential term $U_{\text{NL}}$ is defined as follows:
\begin{equation}
\begin{aligned}
    U_{\text{NL}}(\sigma,\omega,\boldsymbol{\delta},\boldsymbol{\rho}) &= \frac{1}{3}g_2 \sigma^3 + \frac{1}{4}g_3 \sigma^4 -\frac{1}{4}c_3(\omega_\mu \omega^\mu)^2 \\
    &\quad- \Lambda_{\sigma\delta}\sigma^2\boldsymbol{\delta}^2 - \Lambda_{\omega \rho} (\omega_\mu \omega^\mu) (\boldsymbol{\rho}_\nu \cdot \boldsymbol{\rho}^\nu).  
\end{aligned}
\end{equation}
Here, $g_2,g_3, \text{and} \ c_3$ represent the coupling constants of the corresponding mesons, while $\Lambda_{\sigma\delta}$ and $\Lambda_{\omega\rho}$ denote the mixing parameters. In the RMF approximation, the meson fields are replaced by their mean-field expectation values. The effective nucleon mass in the medium is defined as
\begin{equation}
    M^*_N=M_N-g_\sigma \bar{\sigma} \mp g_\delta \bar{\delta}.
\end{equation}
Regarding the sign of the third term, the negative (positive) sign corresponds to contributions to the proton (neutron) mass.
For uniform infinite nuclear matter, the Euler-Lagrange equations derived from the Lagrangian density are reduced to
\begin{subequations}
    \begin{align}
        &\left[ \gamma_\mu (i\partial^\mu  - g_\omega \bar{\omega} \mp g_\rho \bar{\rho} ) - M^*_N  \right] \psi_N = 0, \\
        &m^2_\sigma \bar{\sigma} =g_\sigma(\rho^s_p+\rho^s_n) - g_2 \bar{\sigma}^2 -g_3 \bar{\sigma}^3 + 2\Lambda_{\sigma \delta} \bar{\sigma} \delta^2 ,\\
        &m^2_\omega \bar{\omega} =g_\omega(\rho_p+\rho_n) - c_3\bar{\omega}^3 - 2\Lambda_{\omega \rho} \bar{\omega} \bar{\rho}^2,\\
        &m^2_\delta \bar{\delta} =g_\omega(\rho^s_p-\rho^s_n) + 2\Lambda_{\sigma \delta} \bar{\sigma}^2 \bar{\delta},\\
        &m^2_\rho \bar{\rho} =g_\rho(\rho_p-\rho_n) -2\Lambda_{\omega \rho} \bar{\omega}^2 \bar{\rho}.
    \end{align}
\end{subequations}\noindent
In these expressions, $\rho^s_N$ and $\rho_N$ represent, respectively, the scalar and baryon densities for nucleon species $N=\{p,n\}$. All meson mean fields are obtained in a self-consistent manner by iteratively solving these coupled equations.

The total energy density and pressure of the system is then given by:
\begin{equation}
    \begin{aligned}
        \varepsilon_B &= \frac{1}{\pi^2} \sum_{N} \int^{k_{F_N}}_{0} \dd k k^2 \sqrt{k^2+M^{*2}_N} \\
        &\quad + \frac{1}{2}(m^2_\sigma \bar{\sigma}^2 + m^2_\omega \bar{\omega}^2 + m^2_\delta \bar{\delta}^2 + m^2_\rho \bar{\rho}^2) \\
        &\quad  +\frac{1}{3}g_2 \bar{\sigma}^3 + \frac{1}{4}g_3 \bar{\sigma}^4 + \frac{3}{4} c_3 \bar{\omega^4} \\ 
        &\quad - \Lambda_{\sigma \delta} \bar{\sigma}^2 \bar{\delta}^2 + 3\Lambda_{\omega 
        \rho} \bar{\omega}^2 \bar{\rho}^2,
\end{aligned}
\end{equation}
and
\begin{equation}
    \begin{aligned}
        P_B &= \frac{1}{3\pi^2} \sum_{N}\int^{k_{F_N}}_{0} \dd k \frac{k^4}{\sqrt{k^2+M^{*2}_N}}  \\
        &\quad - \frac{1}{2}(m^2_\sigma \bar{\sigma}^2- m^2_\omega \bar{\omega}^2 + m^2_\delta \bar{\delta}^2 - m^2_\rho \bar{\rho}^2) \\
        &\quad  -\frac{1}{3}g_2 \bar{\sigma}^3 - \frac{1}{4}g_3 \bar{\sigma}^4 + \frac{1}{4} c_3 \bar{\omega^4} \\ 
        &\quad + \Lambda_{\sigma \delta} \bar{\sigma}^2 \bar{\delta}^2 + \Lambda_{\omega 
        \rho} \bar{\omega}^2 \bar{\rho}^2.
\end{aligned}
\end{equation}

Using the proton fraction $Y_p$ and the energy per nucleon in symmetric nuclear matter (SNM) with $Y_p=1/2$, $E_\text{SNM}$, nuclear matter properties are characterized by
\begin{subequations}
    \begin{align}
        K_0 &= 9n_0^2 \left( \frac{\partial^2 E_{\text{SNM}}(n)}{\partial n^2} \right)_{n_0} \quad \text{(incompressibility)}, \\
        S &= \frac{1}{8} \left( \frac{\partial^2 E}{\partial Y_p^2} \right)_{Y_p=1/2} \quad (\text{symmetry energy}),\\
        L &= 3n_0 \left( \frac{\partial S(n)}{\partial n} \right)_{n_0} \quad \text{(slope of $S$)}.
    \end{align}
\end{subequations}

\begin{table}[t]
\centering
\caption{Nuclear matter properties at saturation density for OMEG parameter sets used in the present study. The saturation density $n_0$ is in the units of $\text{fm}^{-3}$, and all the other quantities are given in MeV.}
{\normalsize
\begin{tabular}{lcccccc}
\hline\hline
\textbf{Model} & \boldmath$n_0$ & \boldmath$M^*_N / M_N$ & \boldmath$K_0$ & \boldmath$S_0$  & \boldmath$L$   \\
\midrule
OMEG1 & 0.1484 & 0.620 & 256.0 & 35.06 & 70.00  \\
OMEG2  & 0.1484 & 0.620 & 256.0 & 33.00 & 45.00  \\
OMEG3  & 0.1484 & 0.620 & 256.0 & 30.00 & 20.00  \\
\hline\hline
\end{tabular}
}
\label{tab:nuclear_properties}
\end{table}

For the coupling constants, we employ the OMEG series EoSs which have been constructed in \citet{Miyatsu2023}. Their nuclear matter properties are summarized in Table \ref{tab:nuclear_properties}. From the table, we can find that only $L$ varies, whereas the other properties are almost the same with minor difference in the $S_0$ value.
Thereby we can evaluate the pure effect of $L$ on neutron star properties.

\begin{figure}[h]
    \centering
    \includegraphics[width=0.96\linewidth]{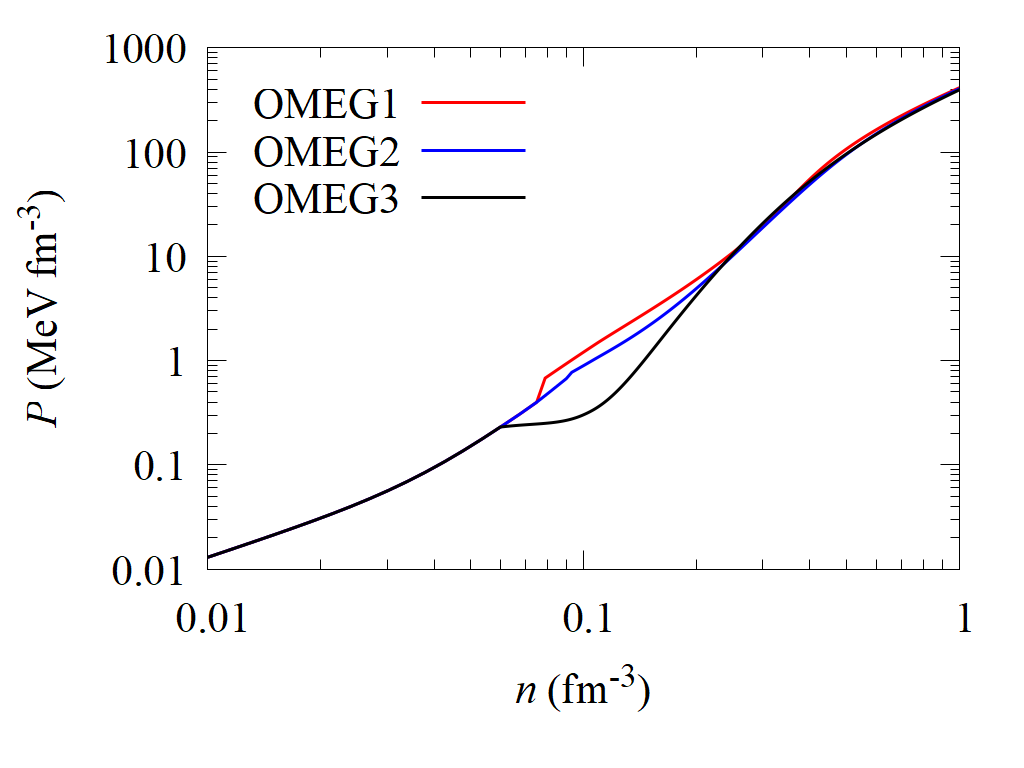}
    \caption{Equations of state calculated under the $\beta$-equilibrium and charge neutrality conditions using OMEG parameter sets. Pressure $P$ is shown as a function of baryon number density $n$. Results are shown for three OMEG parameter sets: OMEG1 (red), OMEG2 (blue), OMEG3 (black)}
    \label{EOS}
\end{figure}

For each OMEG parameter set, Fig.~\ref{EOS} shows the neutron star EoS calculated under the $\beta$-equilibrium and charge neutrality $npe\mu$ conditions, smoothly matched to the MYN13 crust EoS \citep{Miyatsu2013,Miyatsu2024} at low densities. In this process, since the orignial OMEG3 EoS exhibits a region where the core pressure temporarily becomes lower than that of the MYN13 crust, the two segments were artificially connected to ensure continuity.

\subsection{Neutron Star Structure}

The structure of a neutron star can be well described with the general relativity. Within the framework of the general relativity, the spacetime geometry is governed by the Einstein field equations,

\begin{equation}
    G_{\mu\nu} = R_{\mu\nu} - \frac{1}{2} g_{\mu\nu} R = \frac{8\pi G}{c^4} T_{\mu\nu},
    \label{Einstien}
\end{equation}
where $G_{\mu\nu}$ is the Einstein tensor, $R_{\mu\nu}$ the Ricci tensor, $R$ the scalar curvature, and $T_{\mu\nu}$ the stress-energy tensor. We adopt the metric signature $(-, +, +, +)$ and a $c = G = 1$ unit for general relativistic formulation given below. In the case of a perfect fluid, the energy-momentum tensor takes the form
\begin{equation}
    T^{\mu \nu} = (\varepsilon + P)u^{\mu}u^{\nu} + Pg^{\mu \nu},
    \label{EMtensor}
\end{equation}
where $\varepsilon$ and $P$ denote the energy density and pressure, respectively, and $u^\mu$ is the four-velocity of the fluid. These relations provide the foundation for modeling both static and rotating neutron star configurations.

\subsubsection{TOV Equation}

In the non-rotating limit, a neutron star can be modeled as a static, spherically-symmetric configuration. The line element of the spacetime is expressed as
\begin{equation}
    \dd s^2=-e^{2\nu(r)}\dd t^2+e^{2\lambda(r)}\dd r^2+r^2(\dd \theta^2+\sin^2\theta\dd\phi^2),
\end{equation}
where $\nu(r)$ and $\lambda(r)$ are metric potentials describing the gravitational field. Substituting this metric into the Eq.~\eqref{Einstien} and Eq.~\eqref{EMtensor} with energy-momentum conservation $(\nabla_\mu T^{\mu \nu}=0)$, one obtains the TOV equations,
\begin{subequations}
    \begin{align}
        \frac{\dd P}{\dd r} &= - (\varepsilon+P)\frac{M+4\pi r^3P}{r(r-2M)},\\
        \frac{\dd M}{\dd r} &= 4\pi r^2 \varepsilon,\\
        \frac{\dd\nu}{\dd r} &= -2(\varepsilon+P)^{-1} \frac{\dd P}{\dd r},
    \end{align}
\end{subequations}\noindent
as the hydrostatic equilibrium condition. Starting from a central density $\varepsilon_c = \varepsilon(r\!=\!0)$, these equations are integrated outward to determine pressure $P(r)$, energy density $\varepsilon(r)$, and enclosed mass $M(r)$ along $r$ coordinate until the internal pressure vanishes at the surface of the neutron star, $P(r\!=\!R)=0$.

\subsubsection{Hartle-Thorne Perturbative Method}

The HT method provides a perturbative framework for describing slowly rotating neutron stars by expanding the spacetime metric in powers of the angular velocity $\Omega$. Up to the second order in $\Omega$, the metric takes the form
\begin{equation}
    \begin{aligned}
        \dd s^2 &= -e^{\nu}\qty[1 + 2(h_0 + h_2P_2)]\dd t^2 + \\
    &\hspace{5mm} + \frac{1 + 2(m_0 + m_2P_2) / (r - 2M)}{1 - 2M/r}\dd r^2\\
    &\hspace{5mm} + r^2\qty[1+2(v_2-h_2)P_2]\qty[\dd\theta^2 + \sin^2\theta(\dd\phi - \omega \dd t)^2] \\
    &\hspace{5mm}+ \mathcal{O}(\Omega^3),
    \end{aligned}
\end{equation}
where $P_2 = P_2(\cos\theta) = 1/2(3\cos^2\theta -1)$ is the Legendre polynomial.
All quantities, $\nu, h_l, m_l, v_l$ and $\Omega$, depend on the radial coordinate $r$.

\paragraph{First-order equations}
At the first order in $\Omega$, rotation introduces a frame-dragging effect, described by the function $\omega(r)$. 
The angular velocity relative to the local inertial frame, $\bar{\omega}(r) = \Omega - \omega(r)$, is supposed to satisfy the following differential equation
\begin{equation}
    \frac{1}{r^4}\frac{\dd}{\dd r}\qty(r^4 j(r)\frac{\dd\bar{\omega}}{\dd r}) + \frac{4}{r}\frac{\dd j}{\dd r}\bar{\omega} = 0
\end{equation}
with
\begin{equation}
    j(r) = e^{-\nu(r)/2}\qty[1 - 2M(r)/r]^{1/2}.
\end{equation}
Outside the star it has the following form:
\begin{equation}
    \bar{\omega}(r) = \Omega - \frac{2J}{r^3},
\end{equation}
where $J$ is the total angular momentum of the star. Here, $J$ and $\Omega$ are given by
\begin{equation}
    J = \frac{1}{6}R^4\qty(\frac{d\bar{\omega}}{dr})_{r=R},\quad \Omega = \bar{\omega}(R) + \frac{2J}{r^3}.
\end{equation}
In practical calculations, calculated $\Omega$ should coincide with the intended value $\Omega_{\text{fix}}$. To achieve this, we adjust $\bar{\omega}$ according to
\begin{equation}
    \bar{\omega}(r) \to \bar{\omega}(r)\times \frac{\Omega_\text{fix}}{\Omega_\text{calc}}.
\end{equation}
This first-order treatment captures the frame-dragging behavior of the stellar interior and determines the total angular momentum within the slow-rotation approximation.

\paragraph{Second-order equations}

The second order perturbations affect both the spherically-symmetric components $(h_0,m_0)$ and the quadrupole deformations $(h_2,v_2)$ of the stellar structure. The relevant equations describing these quantities are given by
\begin{equation}
\begin{aligned}
    \frac{\dd m_0}{\dd r} &=4\pi r^2 \frac{\dd\varepsilon}{\dd P}(\varepsilon+P)p_0^* + \frac{1}{12}j^2r^4 \left( \frac{\dd\bar{\omega}}{\dd r} \right)^2 \\
    & \quad - \frac{1}{3}r^3 \frac{\dd j^2}{\dd r}\bar{\omega}^2,
\end{aligned}    
\end{equation}
\begin{equation}
\begin{aligned}
    \frac{\dd p_0^*}{\dd r} &=-\frac{m_0 (1+8\pi r^2 P)}{r-2M}-\frac{4\pi (\varepsilon+P)r^2}{(r-2M)^2}p^*_0 \\
    & \quad +\frac{1}{12}\frac{r^4j^2}{(r-2M)} \left( \frac{\dd\bar{\omega}}{\dd r} \right)^2 + \frac{1}{3} \frac{\dd}{\dd r} \left( \frac{r^3 j^2 \bar{\omega}^2}{r-2M} \right),
\end{aligned}
\end{equation}
\begin{equation}
\begin{aligned}
    \frac{\dd v_2}{\dd r} &=-\frac{\dd\nu}{\dd r}h_2 + \left( \frac{1}{r}+ \frac{1}{2} \frac{\dd\nu}{\dd r} \right) \\
    &\quad \left[ -\frac{1}{3}r^3 \frac{\dd j^2}{\dd r}\bar{\omega}^2 + \frac{1}{6}j^2 r^4 \left( \frac{\dd\bar{\omega}}{\dd r}^2 \right) \right],
\end{aligned}
\end{equation}
\begin{equation}
\begin{aligned}
    \frac{\dd h_2}{\dd r} &= \left\{ - \frac{\dd\nu}{\dd r} + \frac{r}{r-2M}\left( \frac{\dd\nu}{\dd r} \right)^{-1} \left[ 8 \pi (\varepsilon+P) - \frac{4M}{r^3}\right] \right\}h_2 \\
    & \quad -\frac{4v_2}{r(r-2M)} \left( \frac{\dd\nu}{\dd r} \right)^{-1} \\
    & \quad+\frac{1}{6} \left[ \frac{1}{2} \frac{\dd\nu}{\dd r} r - \frac{1}{r-2M} \left( \frac{\dd\nu}{\dd r} \right)^{-1} \right]r^3j^2\left( \frac{\dd\bar{\omega}}{\dd r} \right)^2 \\
    &\quad -\frac{1}{3} \left[ \frac{1}{2} \frac{\dd\nu}{\dd r}r + \frac{1}{r-2M}\left( \frac{\dd\nu}{\dd r} \right)^{-1} \right] r^2 \frac{\dd j^2}{\dd r}\bar{\omega}^2.
\end{aligned}
\end{equation}
These coupled equations are integrated from the stellar center to the surface in the same manner as TOV equations. The second-order correction to the pressure perturbation reads
\begin{equation}
    p^*_2 = -h_2 -\frac{1}{3}r^2 e^{-\nu}\bar{\omega}^2.
\end{equation}
The total gravitational mass and radius of the rotating configuration are then given by
\begin{equation}
\begin{aligned}
    M^\text{HT}&=M(r)+m_0(R) +J^2/R^3, \\
    R^\text{HT}&=r+\xi_0 (r) + \xi_2(r)P_2(\cos \theta), \label{eq:HTradius}
\end{aligned}
\end{equation}
where the radial displacements $\xi_0$ and $\xi_2$ are defined as
\begin{equation}
\begin{aligned}
    \xi_0 &= -p^*_0 (\varepsilon+P) / (\dd P/\dd r), \\
    \xi_2 &= -p^*_2 (\varepsilon+P) / (\dd P/\dd r).
\end{aligned}
\end{equation}
These terms, $\xi_0$ and $\xi_2$, represent the radial expansion and quadrupole deformation of the stellar surface induced by rotation, respectively.
They correspond to the cases of $\cos\theta=1$ and $\cos\theta=-1/2$ in Eq.~\eqref{eq:HTradius}, respectively, we thus have
\begin{equation}
    \begin{aligned}
        R_\text{pol} &= r + \xi_0(r) + \xi_2(r),\\
        R_\text{eq} &= r + \xi_0(r) - \frac{1}{2}\xi_2(r).
    \end{aligned}
\end{equation}
where the $R_\text{pol}$ and $R_\text{eq}$ are the radii at the polar point and the equator, respectively.

\subsubsection{Komatsu-Eriguchi-Hachisu Method}

To model rapidly rotating neutron stars in a fully general relativistic manner, one should solve the equations for axially-symmetric hydrostatic equilibrium. In this framework, the spacetime line element is written as
\begin{equation}
    \begin{aligned}
        \dd s^2 &= -e^{\hat{\gamma}+\hat{\varrho}}\dd \hat{t}^2 + e^{2 \hat{\alpha}}(\dd \hat{r}^2 + \hat{r}^2\dd \hat{\theta}^2) \\
        & \quad + e^{\hat{\gamma}-\hat{\varrho}}\hat{r}^2 \sin^2\hat{\theta}(\dd\hat{\phi}-\hat{\omega}\dd \hat{t})^2.
    \end{aligned}
\end{equation}

In the KEH method, the Einstein field equations \eqref{Einstien} are reduced to a set of coupled elliptic differential equations for the metric potentials:
\begin{subequations}
    \label{KEHEQ}
    \begin{align}
        &\nabla^2[\hat{\varrho} e^{\hat{\gamma}/2}] = S_{\hat{\varrho}}(\hat{r},\hat{\mu}), \\
        &\left( \nabla^2 + \frac{1}{\hat{r}} \partial_{\hat{r}} - \frac{\hat{\mu}}{\hat{r}^2} \partial_{\hat{\mu}} \right) [\hat{\gamma} e^{\hat{\gamma}/2}] = S_{\hat{\gamma}}(\hat{r},\hat{\mu}), \\
        &\left( \nabla^2 + \frac{2}{\hat{r}} \partial_{\hat{r}} - \frac{2\hat{\mu}}{\hat{r}^2} \partial_{\hat{\mu}} \right) [\hat{\omega} e^{(\hat{\gamma}-2\hat{\varrho})/2}] = S_{\hat{\omega}}(\hat{r},\hat{\mu}), \\
        & \frac{\partial \hat{\alpha}}{\partial \hat{\mu}} = S_{\hat{\alpha}}(\hat{r},\hat{\mu}),
    \end{align}
\end{subequations}
\noindent\hspace{-5pt}
where $S_{\hat{\varrho}}$, $S_{\hat{\gamma}}$, $S_{\hat{\omega}}$, and $S_{\hat{\alpha}}$ are source terms depending on the local energy density, pressure and angular velocity profiles (see \cite{Komatsu1989a} for detailed expressions). The conservation condition of energy-momentum \eqref{EMtensor} provides the hydrostatic equilibrium condition $(\nabla_{\mu} T^{\mu\nu} = 0)$
\begin{equation}
    \ln{H} + \frac{\hat{\varrho}+\hat{\gamma}}{2} + \frac{1}{2}\ln{(1-v^2)} + \int j(\Omega)\dd\Omega = C,
    \label{eqcon}
\end{equation}
with
\begin{equation}
    \ln{H} = \int \frac{\dd P}{\varepsilon + P} .
\end{equation}
where $H$ is the enthalpy, $v$ is the fluid velocity measured in the local inertial frame, and $C$ is an integration constant. For calculating a rotating neutron star, one of the most widely used models is the one-parameter law:
\begin{equation}
    j(\Omega) = A^2(\Omega-\Omega_c).
\end{equation}
In this study, we assume the rigid rotation, which corresponds to the limit $A \rightarrow \infty$. Therefore, the contribution from the differential rotation is neglected here. Each metric potential \eqref{KEHEQ} can be expressed as an integral form of Poisson-like equations, and the field equations together with the equilibrium condition \eqref{eqcon} are solved iteratively until full self-consistency is achieved.

\begin{figure}[t]
    \centering
    \includegraphics[width=0.96\linewidth]{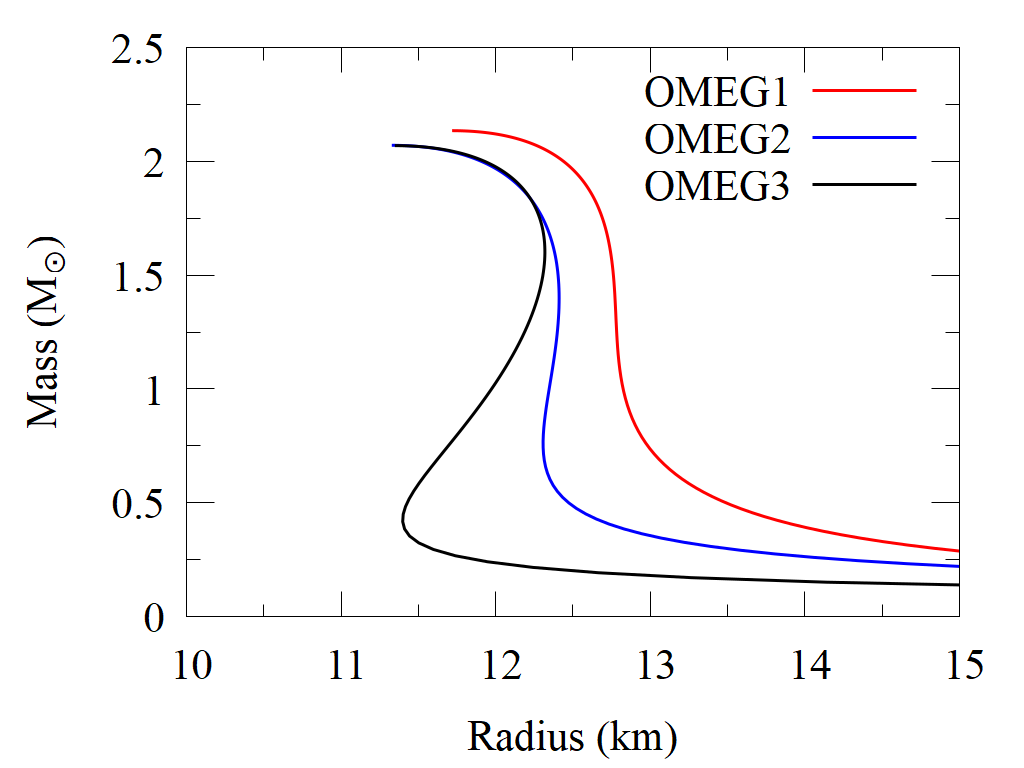}
    \caption{Mass-radius relations for neutron stars obtained from the TOV equation using the OMEG+MYN13 EoSs. The red, blue, and black curves correspond to OMEG1, OMEG2, and OMEG3, respectively.}
    \label{TOVMR}
\end{figure}

\begin{figure}[t]
    \centering\vspace{-5mm}
    \includegraphics[width=0.96\linewidth]{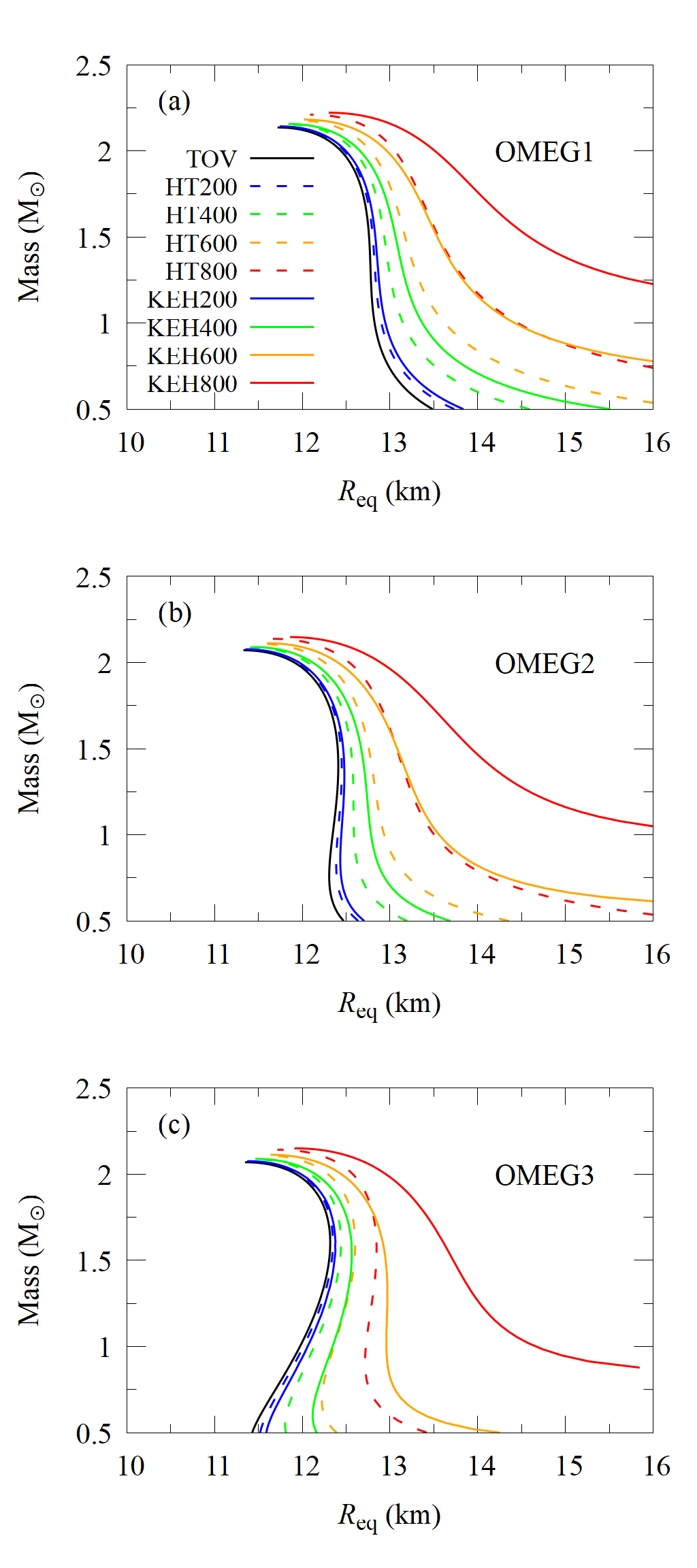}
    \caption{Mass–radius relations for neutron stars calculated using the HT (dashed lines) and KEH (solid lines) methods at various spin frequencies (200, 400, 600, and 800 Hz) for three different EoSs: OMEG1 (a), OMEG2 (b), and OMEG3 (c). The black solid line represents the TOV result. Both methods show that the neutron star mass and radius increase with spin frequency, and the difference between the HT and KEH results becomes significant above ~400 Hz, especially for a canonical $1.4 \ M_\odot$ neutron star.}
    \label{RotatingMR}
\end{figure}

Once the metric potentials are determined by the self-consistent field method, the gravitational mass and total angular momentum are calculated through the following relation:
\begin{equation}
\begin{aligned}
M &= 2\pi \iint \dd \hat{r}\,\dd\hat{\theta}\, \hat{r}^2 \sin{\hat{\theta}}\, e^{2\hat{\alpha}+\hat{\gamma}} \\
&\quad \times \Biggl[
\frac{(\varepsilon+P)(1+v^2)}{1-v^2} \\
&\qquad + 2P
+ 2\hat{r}\sin{\hat{\theta}}\, \hat{\omega} e^{-\hat{\varrho}} \frac{(\varepsilon+P)v}{1-v^2}
\Biggr].
\end{aligned}
\end{equation}

\begin{equation}
    J = 2\pi \iint \dd \hat{r}\,\dd \hat{\theta} \, \hat{r}^3 \sin^2{\hat{\theta}}\, e^{2\hat{\alpha}+\hat{\gamma}-\hat{\varrho}} \frac{(\varepsilon+P)v}{1-v^2}.
\end{equation}
The circumferential radius at the equator is defined by

\begin{equation}
    R_\text{eq}=\hat{r}_e e^{(\hat{\gamma}_e + \hat{\varrho}_e)/2}
\end{equation}

This calculation is performed by inputting the polar-to- radius ratio, $\hat{r}_\text{ratio}=\hat{r}_{\text{p}}/\hat{r}_{\text{e}}$. The angular velocity $\Omega$, and the radius $\hat{r}_{e}$ are determined after carrying out the self-consistent calculation. In this work, we adopted the CST approach, which maps the KEH calculation onto a finite coordinate domain.

\section{Numerical Results}

We perform calculations based on both the HT and KEH methods using the OMEG EoSs. The initial conditions for both methods are derived from the TOV equation. 
Figure~\ref{TOVMR} shows the mass-radius relation obtained from the TOV calculations with the examined three parameter sets.
From this figure, it can be seen that EoSs with higher $L$ values yield larger radii for a $1.4 \ \text{M}_\odot$ neutron star, in the order of OMEG 1, 2, and 3.

In asymptotically flat spacetimes, the leading-order terms in the metric expansion correspond to the mass and angular momentum, which are coordinate-invariant global charges defined at spatial infinity. Since these quantities are directly related to observable properties of compact stars, a meaningful comparison between different general relativistic models should be performed at fixed values of $M$ and $J$.
As shown in Table~2, for a $1.4\,M_\odot$ model with $J = 0.3$, the HT and KEH methods yield nearly identical values for the central density,  radius, and spin frequency. This indicates that, in the slow-rotation regime, the two approaches describe essentially the same stellar configuration.
In contrast, for $J = 1.5$, pronounced differences emerge between the HT and KEH results. This behavior reflects the limitation of the perturbative expansion underlying the HT formalism at higher rotational frequencies, where higher-order rotational effects become significant. Therefore, this limitation must be carefully considered when interpreting the numerical results. However, in this study we use frequency $f$ for the simplicity of comparisons.

\begin{table}[t]
\centering
\setlength{\tabcolsep}{8pt}
\setlength{\textfloatsep}{10pt}
\caption{Neutron star properties for fixed angular momentum $J = 0.3$ and $J = 1.5$. Shown are the central density $\rho_c$, the  equatorial radius $R_{\rm eq}$, and the rotation frequency $f=\Omega/2\pi$ at $1.4M_{\odot}$ for the HT and KEH methods. 
}   

{\normalsize
\begin{tabular}{ccccc}
\hline\hline

\multicolumn{5}{c}{$J = 0.3$ $\text{km}^2$} \\
\hline
\multicolumn{2}{c}{Model} & $\rho_c$ ($g/\text{cm}^3$)& $R_{\rm eq}$ (\text{km}) & $f$ (\text{Hz}) \\
\hline
\multirow{2}{*}{OMEG1} & HT & 0.72$\times10^{15}$ & 12.79 & 123 \\
                       & KEH & 0.72$\times10^{15}$ & 12.80 & 123 \\
\multirow{2}{*}{OMEG2} & HT & 0.77$\times10^{15}$ & 12.42 & 126 \\
                       & KEH & 0.77$\times10^{15}$ & 12.44 & 126 \\
\multirow{2}{*}{OMEG3} & HT & 0.73$\times10^{15}$ & 12.28 & 124 \\
                       & KEH & 0.73$\times10^{15}$ & 12.30 & 123  \\

\hline\hline
\multicolumn{5}{c}{$J = 1.5$ $\text{km}^2$} \\
\hline
\multicolumn{2}{c}{Model} & $\rho_c$ ($g/\text{cm}^3$) & $R_{\rm eq}$ (\text{km}) & $f$ (\text{Hz}) \\
\hline
\multirow{2}{*}{OMEG1} & HT & 0.67$\times10^{15}$ & 13.44 & 709 \\
                       & KEH & 0.69$\times10^{15}$ & 13.53 & 569 \\
\multirow{2}{*}{OMEG2} & HT & 0.71$\times10^{15}$ & 13.00 & 730 \\
                       & KEH & 0.73$\times10^{15}$ & 13.11 & 585 \\
\multirow{2}{*}{OMEG3} & HT & 0.67$\times10^{15}$ & 12.73 & 733 \\
                       & KEH & 0.70$\times10^{15}$ & 12.90 & 575  \\

\hline\hline
\end{tabular}
}
\label{tab:J_models}
\end{table}

\begin{deluxetable*}{cccccccccc}
\setlength{\tabcolsep}{5pt} 
\setlength{\textfloatsep}{10pt} 
\tablewidth{0pt}
\caption{Comparison of maximum mass ($M_{\text{max}}$), corresponding radii ($R_{\text{max}}$ and $R_{1.4}$) and angular momentum ($J_{\text{1.4}}$) 
for three EoSs (OMEG1–OMEG3) at different spin frequencies $f = 800$, 200, and 0~Hz. 
The $f = 0$ rows represent non-rotating configurations. 
Results are shown for both KEH and HT methods. \label{tab:description}}
{\normalsize
\tablehead{
\colhead{Model} & \colhead{$f$ (Hz)} & \colhead{$M^{\text{KEH}}_{\text{max}}$ $(M_\odot)$} & \colhead{$M^{\text{HT}}_{\text{max}}$ $(M_\odot)$} & \colhead{$R^{\text{KEH}}_{\text{max}}$ (km)} & \colhead{$R^{\text{HT}}_{\text{max}}$ (km)} & \colhead{$R^{\text{KEH}}_{\text{1.4}}$ (km)}  & \colhead{$R^{\text{HT}}_{\text{1.4}}$ (km)} &
\colhead{$J^{\text{KEH}}_{\text{1.4}}$ ($\text{km}^2$)}  & \colhead{$J^{\text{HT}}_{\text{1.4}}$ ($\text{km}^2$)}
}
\startdata
\multirow{3}{*}{OMEG1} & 800 & 2.22 & 2.21 & 12.30 & 12.09 & 14.91 & 13.65 & 2.36 & 1.62 \\
                       & 200 & 2.14 & 2.14 & 11.74 & 11.73 & 12.85 & 12.82 & 0.49 & 0.49  \\
                       & 0 & \multicolumn{2}{c}{2.14}  & \multicolumn{2}{c}{11.71} & \multicolumn{2}{c}{12.77} & \multicolumn{2}{c}{0} \\
\hline
\multirow{3}{*}{OMEG2} & 800 & 2.15 & 2.14 & 11.86 & 11.67 & 14.12 & 13.14 & 2.27 & 1.59 \\
                       & 200 & 2.08 & 2.08 & 11.35 & 11.34 & 12.48 & 12.45 & 0.48 & 0.47\\
                       & 0 & \multicolumn{2}{c}{2.07}  & \multicolumn{2}{c}{11.32} & \multicolumn{2}{c}{12.41} & \multicolumn{2}{c}{0} \\
\hline
\multirow{3}{*}{OMEG3} & 800 & 2.15 & 2.14 & 11.91 & 11.72 & 13.82 & 12.83 & 2.30 & 1.58 \\
                       & 200 & 2.08 & 2.07 & 11.37 & 11.36 & 12.34 & 12.30 & 0.49 & 0.48 \\
                       & 0 & \multicolumn{2}{c}{2.07}  & \multicolumn{2}{c}{11.34} & \multicolumn{2}{c}{12.27} & \multicolumn{2}{c}{0} \\\hline\hline
\enddata
}
\end{deluxetable*}

Figure \ref{RotatingMR} compares the mass-radius relations obtained from the HT (dashed lines) and KEH (solid lines) methods at different spin frequencies. In both methods, the neutron star radius and mass increase with higher spin frequency. In particular, above 400 Hz, both calculations deviate noticeably from the non-rotating TOV results, showing that the radius of a $1.4 \ M_\odot$ neutron star increases on the order of kilometers. Among the three parameter sets under study, the OMEG1 case with the largest $L=70$ MeV [Fig.~\ref{RotatingMR} (a)] shows the largest increase in radius due to rotation, followed by OMEG2 [Fig.~\ref{RotatingMR} (b)] and OMEG3 [Fig.~\ref{RotatingMR} (c)], for both HT and KEH methods. This indicates that not only the radius itself, but also the enlargement of the radius due to rotation, correlates with the value of $L$. This trend is consistent with previous studies by \cite{Lopes2024} and \cite{Kwon2025}. Since all other nuclear matter properties are fixed while only the $L$ value varies in the OMEG EoSs, we can attribute this increase in neutron star radius mainly to the effect of the $L$ parameter.

When comparing the HT and KEH methods, the deviation between the two methods depends on the position along the mass-radius curves. For neutron stars at the maximum mass configuration, the increase in both mass and radius are small, and the difference between the two methods remains minor. Even for the case of $f=800$ Hz, the differences of masses and radii at maximum mass are limited within about 1\% and  2\%, respectively. On the other hand, for a typical $1.4M_\odot$ neutron star, the radius increase is substantial, especially for high frequency cases.
While the difference is only around 0.2--0.3\% for the 200 Hz case, the 800 Hz frequency provides the 6.9--8.4\% discrepancy of radius between the two methods. As shown in Table~3, $J$ obtained from the two methods are in good agreement for slowly rotating models (200~Hz). 
However, at higher spin frequencies, a significant discrepancy in the $J$ appears, even though the masses remain similar in both calculations. This indicates that the perturbative HT method becomes inaccurate in the high frequency regions and fully general relativistic treatments such as the KEH method are required.

\begin{figure}[h]
    \centering
    \includegraphics[width=0.96\linewidth]{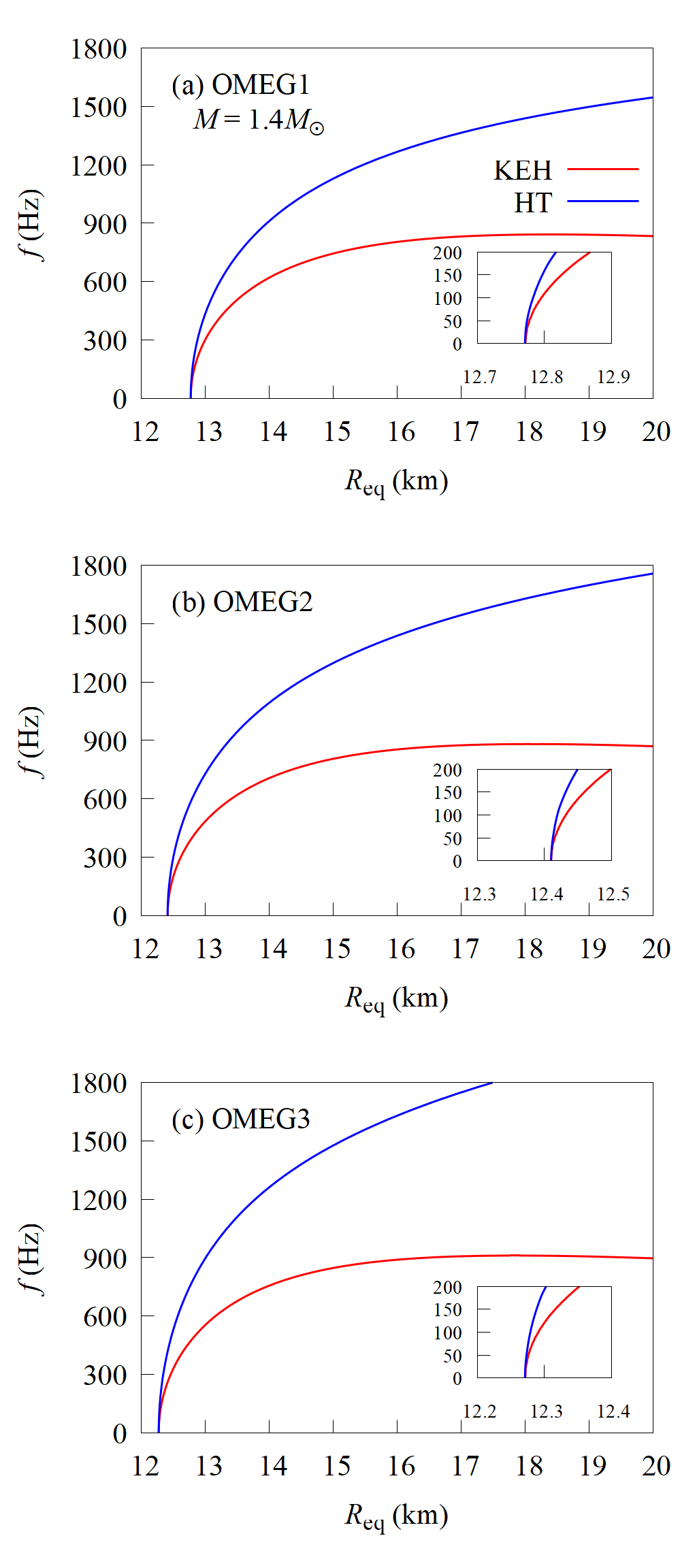}
    \caption{spin frequency $(f)$ is shown as a function of the  radius, $R_{\text{eq}}$, for neutron stars with $M=1.4M_\odot$. Panels (a), (b), and (c) correspond to the OMEG1, OMEG2, and OMEG3 EOSs, respectively. The red and blue curves represent results from the KEH and HT methods, respectively. The inset panel shows an enlarged view of the nearly spherical regime, where the two methods yield nearly identical results below 200 Hz.}
    \label{OMEGOMGEA}
\end{figure}

\begin{table}[b]
\centering
\setlength{\tabcolsep}{8pt} 
\setlength{\textfloatsep}{10pt} 
\caption{Comparison of  radius ($R_{\text{eq}}^{\text{HT}}$ and $R_{\text{eq}}^{\text{KEH}}$) 
for three rotating models (OMEG1–OMEG3) at different spin frequencies $f = 800$, 200, 100, and 10~Hz. 
For the ``Diff'' column values are calculated as
$\text{(Diff)} = (R^{\text{KEH}}_{\text{eq}} -R^{\text{HT}}_{\text{eq}})/R^{\text{KEH}}_{\text{eq}} \times 100$.
The last column indicates the relative error between the two formalisms, showing that both results agree within a few percent for $f\lesssim200$ Hz.}
{\normalsize
\begin{tabular}{ccccc}
\hline\hline
Model & $f \ (\text{Hz})$ & $R^{\text{HT}}_{\text{eq}}$ & $R^{\text{KEH}}_{\text{eq}}$ & Diff (\%)\\
\hline
\multirow{3}{*}{OMEG1} 
 & 800 & 13.659 & 15.906 & 14.1 \\
 & 200 & 12.817 & 12.868 & 0.4 \\
 & 100 & 12.783 & 12.795 & 0.1 \\
 & 10  & 12.771 & 12.771 & $\approx$ 0 \\
\hline
\multirow{3}{*}{OMEG2} 
 & 800 & 13.140 & 14.919 & 11.9 \\
 & 200 & 12.450 & 12.500 & 0.4 \\
 & 100 & 12.419 & 12.432 & 0.1 \\
 & 10  & 12.410 & 12.410 & $\approx$ 0 \\
\hline
\multirow{3}{*}{OMEG3} 
 & 800 & 12.831 & 14.390 & 10.8 \\
 & 200 & 12.302 & 12.352 & 0.4 \\
 & 100 & 12.279 & 12.291 & 0.1 \\
 & 10  & 12.271 & 12.271 & $\approx$ 0 \\\hline\hline
\end{tabular}
}
\label{tab:Hzratio}
\end{table}
To perform a more detailed comparison between the HT and KEH methods, in Fig.~\ref{OMEGOMGEA}, we show the relationship between spin frequency $f$ and the  radius $R_\text{eq}$. 
This figure shows that the parameter sets with smaller $L$ values result in higher spin frequencies to achieve the same degree of radius.
For instance, at the point where $R_\text{eq}=15 \ \text{km}$, the corresponding frequency is around 780 Hz for OMEG1 case with the largest slope of the symmetry energy $L$, whereas that is around 850 Hz for OMEG3 with smallest $L$. The same tendency is also observed in the case of the HT method; the $R_\text{eq}=15 \ \text{km}$ corresponds to $f=1130$ Hz for the OMEG1 case, while it takes around 1500 Hz when the OMEG3 is employed.

This behavior indicates that EoSs with larger $L$ values are more sensitive to changes in radii affected by rotation, which is consistent with the mass-radius relations discussed earlier.
This trend agrees well with the results of \cite{Kwon2025}, and the present study reinforces this interpretation by varying only $L$ while keeping all other nuclear matter properties fixed.

Finally, let us discuss the difference between the HT and KEH methods.
Overall, the HT method yields higher spin frequencies than the KEH method for the same degree of radius. For all the three EoSs examined, the difference between the two approaches becomes noticeable above approximately 200 Hz. Additionally, Fig.~\ref{OMEGOMGEA} shows that, even in the very slowly rotating case with frequency around $f=100$ Hz, the corresponding values are slightly different by around $0.1$ \%.
Furthermore, this difference becomes larger and larger as the spin frequency increases.
This behavior becomes more clear in the high $L$ case. Comparing the cases in 800 Hz frequency, the discrepancy between the HT and KEH method is around 11 \% when OMEG2 or OMEG3 is incorporated, while it reaches to 14.1 \% in OMEG1 case.
The obtained results demonstrate that, to discuss the rotational stability of neutron stars highly accurately, it is essential to incorporate the fully general relativistic frameworks, particularly for high spin frequencies.

\section{Summary and Conclusion}

In this study, a comparative study of rotating neutron stars has been carried out employing two prevailing methods; one is the perturbative Hartle-Thorne (HT) approach and the other is the fully relativistic Komatsu-Eriguchi-Hachisu (KEH) method which was employed in our previous work \cite{Kwon2025}.
Additionally, we have investigated how the difference between these two methods depends on the nuclear matter properties, especially the $L$ value.
For this purpose, we have adopted three nuclear equations of state, OMEG1, OMEG2, and OMEG3, in which only the slopes of the symmetry energy $L$ are different and the others are kept fixed. From the calculations of the mass-radius relation, we have found that the two methods examined show the slight difference in radius profiles even in the slowly-rotating case with $f =200$ Hz. This difference reaches around one kilometer when the frequency reaches 800 Hz. Furthermore, we have found that this behavior becomes clearer as the slope of the symmetry energy $L$ increases.

For a more detailed comparison, we have investigated the relationship between spin frequency and the radius $R_\text{eq}$ for a $1.4 \ M_\odot$ neutron star. From the calculations, we have found that the obtained profiles with the KEH and HT methods are different even for the slowly rotating case $f \simeq100$ Hz. As in the case of mass-radius curves, this characteristic is made more pronounced as the slope of the symmetry energy $L$ increases. Consequently, we have concluded that to accurately investigate the rotation effects and stability of rapidly rotating neutron stars, the second-order perturbative method is insufficient, and the fully general relativistic models are necessarily required.

We point out here that the present analysis with the HT method is based on the second-order perturbative expansion, and the use of higher-order expansions, e.g. the third- \citep{Omar2005} or the seventh-order \citep{Carlos2025} ones, should improve the accuracy of the HT method, although it requires heroic effort to implement complicated equations. 

In addition, the OMEG series employed in this study are intended to compare the dependence of the rotational effects on the slope of the symmetry energy, the $L$ parameter. It can also be considered as our future work to investigate the effects of other nuclear matter properties, such as incompressibility $K$ as well as effective mass $M^*$. Our research aims to comprehensively elucidate the properties of rotating neutron stars based on the fully general relativistic framework. As future work, we plan to analyze the stability of rotating neutron stars and to investigate structural changes that arise when additional components such as differential rotation or exotic particles (hyperons or quarks) in the inner core region are included. 

\begin{acknowledgments}
This work is supported by JST SPRING, Grant Number JPMJSP2180, Sasakawa Scientific Research Grant from the Japan Science Society, as well as Hiki Foundation, Institute of Science Tokyo. One of the authors (K.Y.) acknowledges the financial support as the JSPS Research Fellow, Grant No.~JP24KJ1110. One of the authors (K.S.) is supported by JSPS Grant-in-Aid for Scientific Research, Grants No.\ JP23K03410, No.\ 23K25864, and No.\ JP25H01269. T.M. and M.-K.C. were supported by the Basic Science Research Program through the National Research Foundation of Korea (NRF) under Grant Nos. RS-2025-16071941, RS-2023-00242196, and RS-2021-NR060129.

\end{acknowledgments}

\bibliography{reference}{}
\bibliographystyle{aasjournalv7}

\end{document}